\documentclass[twocolumn,showpacs,prl,superscriptaddress]{revtex4}
\usepackage{amssymb}

\usepackage{amsmath}
\usepackage{graphicx}
\usepackage{dcolumn}
\usepackage{bm}

\begin{document}


\title{Characterization of the initial filamentation of a relativistic electron beam passing through a plasma}

\author{A. Bret}
 \affiliation{ETSI Industriales, Universidad de Castilla-La
Mancha, 13071 Ciudad Real, Spain}

\author{M.-C. Firpo}
 \affiliation{Laboratoire de Physique et Technologie des Plasmas (C.N.R.S. UMR 7648),
Ecole Polytechnique, 91128 Palaiseau cedex, France}

  \author{C. Deutsch} 
\affiliation{Laboratoire de Physique des Gaz et des Plasmas
(C.N.R.S. UMR 8578), Universit\'{e} Paris XI, B\^{a}timent 210,
91405 Orsay cedex,
France}%

\date{\today }

\preprint{}
\begin{abstract}
The linear instability that induces a relativistic electron beam
passing through a return plasma current to filament transversely is
often related to some filamentation mode with wave vector normal to
the beam or confused with Weibel modes. We show that these modes may
not be relevant in this matter and identify the most unstable mode
on the two-stream/filamentation branch as the main trigger for
filamentation. This sets both the characteristic transverse and
longitudinal filamentation scales in the non-resistive initial
stage.
\end{abstract}


\maketitle


Inertial confinement fusion schemes commonly involve in their final
stage the interaction between some highly energetic particle beams
and a dense plasma target. This is in particular valid for the Fast
Ignition Scenario \cite{Tabak} (FIS) where some laser-produced
relativistic electron beam would eventually propagate into the dense
plasma where it would be stopped. This process would lead to strong
local heating and the ignition of a fusion burn wave. In this
respect, microscopic turbulence in beam-plasma systems is one of the
main potentially deleterious effects for inertial fusion schemes
since it may prevent the conditions for burn to be met by broadening
the phase area where particles deposit their energy. Within the FIS
framework, a strong research effort has thus been put recently on
the interaction of a relativistic electron beam with a plasma with a
focus on beam filamentation instability, that is microscopic in the
transverse direction (see e.g.
\cite{Ruhl,Dodd,silva2002,Fonseca2003,Sentoku}). The experimental
evidence of filamentation of very high current laser-produced
electron beams was recently reported for conditions relevant to the
FIS \cite{Tatarakis}. More generally, filamentation is a potential
instability in beam-plasma systems in frameworks ranging from
accelerators physics to solar flares.

\begin{figure}[t]
\begin{center}
\includegraphics[width=0.3\textwidth]{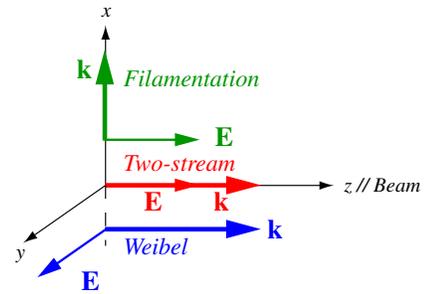}
\end{center}
\caption{Weibel, two-stream and filamentation modes.}
\label{fig:modes}
\end{figure}

In the linear stage, filamentation is generally studied under some
simplifying ab-initio transverse approximation of the dielectric
tensor, so that filamentation instability is attributed to the
exponential growth of unstable electromagnetic purely transverse
modes ($\mathbf{k}\cdot\mathbf{E}=0$) with wave vector $\mathbf{k}$
normal to the beam
\cite{LeeLampe,Hubbard,Okada1980,Okadacollisions,silva2002,Molvig}.
It is also common to refer to this instability as Weibel instability
\cite{LeeLampe,Tatarakis,silva2002},  though the original mode
studied by Weibel \cite{Weibel} would require some plasma
temperature anisotropy to be driven. Figure \ref{fig:modes} sketches
the original definitions of various modes under the original
Weibel's scenario where $\mathbf{k}$ is parallel to the beam, along
the low temperature axis.  As long as the beam is not relativistic,
the largest instability it undergoes is the two-stream one, where
the second ``stream'' is the return current it generates in the
plasma. But in the relativistic regime, the ``filamentation'' growth
rate eventually exceeds the two-stream one and is supposed to induce
beam filamentation.

In reality, the beam suffers much more instabilities at the same
time. Indeed, filamentation, Weibel or two-stream instabilities
pertain to various orientations of the wave vector and various
kinds of waves (transverse or longitudinal), but in the real world
the beam-plasma system triggers every possible modes allowed by
Maxwell equations with a wide range of wave vector orientation.
Among all the triggered modes, the unstable ones shall start
growing exponentially while the most unstable one shall mostly
shape the beam. When it comes to know how the beam is eventually
affected when entering the plasma, one needs therefore to answer
the two questions: 1) which is the most unstable mode all over the
$\mathbf{k}$ space for the system investigated ? and 2) how shall
this mode shape the beam ? Following the guideline built by these
two questions, we assert that the so-called ``filamentation''
instability is not the fastest growing instability, even in the
relativistic regime, so that it is not the answer to the first
question. As for the second question, we shall see that this
instability would not produce the observed effects anyway, even if
it were the stronger one. We shall conclude proposing a new
``candidate'' for beam filamentation and comparing our predictions
with the experimental results presented in \cite{Tatarakis}. For
clarity, we shall keep labelling the most unstable transverse mode
with wave vector normal to the beam as the ``filamentation'' mode,
though our point is precisely that it does not filament.

Let us consider a beam of electrons (having mass $m_{e}$ and
charge $e$) of density $n_b$ and relativistic velocity $V_b$
passing through a return current of plasma electrons of density
$n_p$, so that the system is unmagnetized. Both beam and plasma
are infinite and homogenous and ions are supposed to form a fixed
neutralizing background. Let us define the ratio $\alpha=n_b/n_p$
and introduce the plasma frequency $\omega_{p}=(4\pi
n_{p}e^2/m_{e})^{1/2}$. Here the beam will be assumed to be cold
in the longitudinal direction, which is correct provided the ratio
of its longitudinal thermal velocity $V_{tb\parallel}$ over the
parallel phase velocity $\omega_{p}/k_{\parallel}$ is small
compared to $\alpha^{1/3}$. The filamentation growth rate can then
be evaluated in the weak beam density limit ($\alpha \ll 1$)
through
\begin{equation}\label{eq:tauxfila}
 \delta_F\simeq \beta\sqrt{\frac{\alpha}{\gamma_b}}\omega_p,
\end{equation}
with $\beta=V_b/c$ and $\gamma_b=1/\sqrt{1-\beta^2}$. Within the
same weak beam density limit, the two-stream growth rate reads
\begin{equation}\label{eq:tauxTS}
 \delta_{TS} \simeq
 \frac{\sqrt{3}}{2^{4/3}}\frac{\alpha^{1/3}}{\gamma_b}\omega_p.
\end{equation}
Since $\delta_F$ decreases like $\gamma_b^{-1/2}$ whereas
$\delta_{TS}$ decreases like $\gamma_b^{-1}$, the filamentation
growth rate eventually exceeds the two-stream one when the beam is
relativistic. Comparing filamentation growth rate with the Weibel
one (transverse waves with wave vector along the beam, as in
\cite{Weibel}), one finds filamentation to be also dominant so
that it eventually appears to be the largest instability
\footnote{Transverse beam temperature can reduce it dramatically
\cite{silva2002}, see comments bellow.}.

However, this conclusion  needs to be modified when accounting for
every other unstable modes with wave vector neither normal nor
parallel to the beam. Investigating these modes demands a fully
electromagnetic formalism which is the only way to capture
longitudinal modes (two-stream) as well as transverse modes
(Weibel and filamentation). Indeed, such a procedure shows that
two-stream and filamentation modes pertain to the same branch of
the dispersion equation so that it is possible to switch
continuously from the former to the later by increasing the angle
$\theta_k$ between the beam and the wave vector from 0 to $\pi/2$.
Consequently, the angle $\varphi_k$ between the wave vector and
the electric field of the mode
($\varphi_k=(\widehat{\mathbf{k},\mathbf{E}})$) needs to go
continuously from 0 to $\pi/2$ to bridge between longitudinal
two-stream modes and transverse filamentation modes. In a recent
paper \cite{Bret1}, we began to implement such an electromagnetic
formalism using the relativistic Vlasov equation to describe the
evolution of the electronic distribution function of the
beam-plasma system. Using some simple waterbag distribution
functions for the beam and the plasma, we investigated the
two-stream/filamentation (TSF) branch and found that the growth
rate reaches a maximum for an intermediate orientation of the wave
vector. This maximum scales like $\gamma_b^{-1/3}$ and reads
\begin{equation}\label{eq:tauxmax}
 \delta_{M} \simeq
 \frac{\sqrt{3}}{2^{4/3}}\left(\frac{\alpha}{\gamma_b}\right)^{1/3}\omega_p.
\end{equation}
It is noticeable that this result may be recovered under the
electrostatic longitudinal approximation \cite{Fainberg}. Such an
approach cannot however sweep the whole $\mathbf{k}$-plane.
Equation (\ref{eq:tauxmax}) shows that, even in the relativistic
regime, filamentation growth rate should not be the larger one. On
the contrary, this mixed two-stream filamentation mode shall be
all the more dominant over the usual filamentation mode that the
beam is relativistic because of its $\gamma_b$ scaling. This trend
amplifies even more when accounting for transverse beam
temperature, since filamentation is damped \cite{silva2002,Bret2}
while $\delta_{M}$ is almost unaffected \cite{Bret2}. Therefore,
one can say that the so-called filamentation instability may not
be the fastest growing one.

Let us explore this further and move to our second point by
questioning on what filamentation instability would do to the
beam, if even it had the largest growth rate. Within the linear
approximation, one restricts to small fluctuations of the electron
charge density. If $\rho_1$ and $\mathbf{E}_1$ denote the first
order perturbations, respectively, to the electron charge density
and to the electric field, Poisson equation written in Fourier
space brings $\mathbf{k}\cdot \mathbf{E}_1(\mathbf{k},\omega)=
4\pi \rho_1(\mathbf{k},\omega)$. It comes directly from this
equation that a transverse mode with $\mathbf{k}\cdot
\mathbf{E}_1=0$ has $\rho_1(\mathbf{k},\omega)=0$ and cannot yield
density perturbations within the limits of the linear regime. As a
consequence, the transverse filamentation instability with
wave-vector normal to the beam cannot yield any charge density
fluctuations from the linear stage. It is important to note that,
for the same reasons, the original Weibel mode \cite{Weibel}
cannot linearly induce density filamentation either.

As far as beam filamentation is concerned, experiments and
simulations show that the electronic density varies transversely to
the beam, producing the filaments
\cite{Tatarakis,Ruhl,Sentoku,Fonseca2003}. Now, if the electronic
density varies while background ions are (almost) at rest, there is
necessarily a net charge perturbation which precisely cannot be
accounted for by the mere exponential growth of a purely transverse
wave. It seems therefore that even if it were the fastest growing
instability, the so-called filamentation instability would not
produce these filaments. It is worth noticing that it could produce
\emph{current} filaments, for Maxwell's equations allow such a wave
to produce such perturbations. But these current filaments would
have to preserve the neutrality of the system beam-plasma, that is
to preserve electronic density since ions can be considered at rest.

\begin{figure*}[t]
\begin{center}
\includegraphics[width=.9\textwidth]{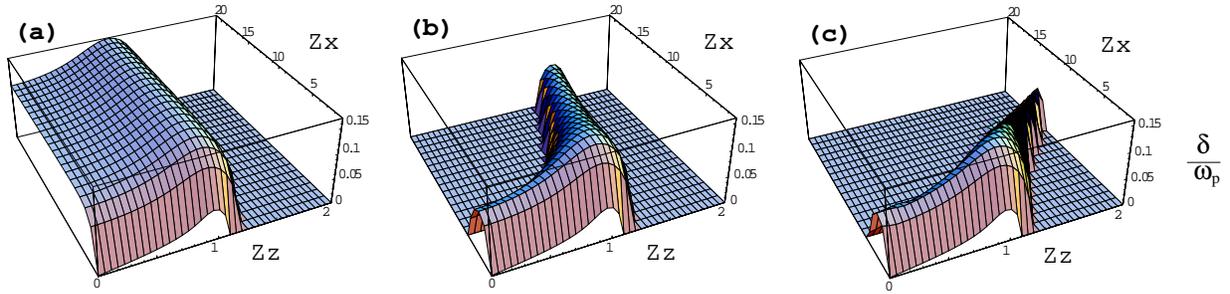}
\end{center}
\caption{Growth rates in terms of
$\mathbf{Z}=\mathbf{k}V_b/\omega_p$. (a) Cold beam - cold plasma
(see also \cite{califano3}), (b) hot  beam - cold plasma and (c)
hot beam - hot plasma. Parameters are $\alpha=0.05$ and
$\gamma_b=4$ for (a,b,c), $V_{tb\perp}=V_b/10$ for (b,c) and
$V_{tp}=V_b/10$ for (c).} \label{fig:growthrates}
\end{figure*}

Let us eventually determine which mode is responsible for the
observed filamentation. We shall see here that the most natural
candidate is the most unstable mode found along the TSF branch.
Being the fastest growing mode, it is the one whose growth should
``shape'' the beam during the linear phase while the other modes
 create fluctuations around this basic shape. As for its
ability to create filaments, it is quasi-longitudinal
\cite{Bret1,Bret2} so that its divergence does not vanish. This
mode, unlike the so-called filamentation mode,  satisfies
therefore the criteria to induce filamentation: It is the fastest
growing one, it is microscopic in the transverse direction and it
is two-stream like, that is, quasi-longitudinal. Expressing the
density perturbation in terms of the wave electric field yields
$\rho_1(\mathbf{k},\omega)=k
E_1(\mathbf{k},\omega)\cos\varphi_k/4\pi$, and one retrieves the
density perturbation in the real space through
\begin{equation}\label{eq:n1fourier}
\rho_1(\mathbf{r},t)= \sum_{\mathbf{k},\omega_{\mathbf{k}}}\frac{k
E_1(\mathbf{k},\omega_{\mathbf{k}})\cos\varphi_k}{4\pi}\exp
(i\mathbf{k}\cdot \mathbf{r}-i\omega_{\mathbf{k}} t)
\end{equation}
The sum above runs over every wave vector and every proper
frequencies $\omega_{\mathbf{k}}$. Yet it will obviously be
dominated by the contribution of the fastest growing (having
$\delta_{\mathbf{k}}\equiv \Im(\omega_{\mathbf{k}})>0$)
self-excited modes. Figure \ref{fig:growthrates} displays the
growth rates on the TSF branch in the $(k_\perp,k_\parallel)$
plane \footnote{Due to the symmetry of the problem, we only
represent the $k_\perp >0$, $k_\parallel>0$ quarter.} for some
zero or finite plasma thermal velocities
$V_{tp}=V_{tp\parallel}=V_{tp\perp}$ and some zero or finite beam
transverse thermal velocities $V_{tb\perp}$ \footnote{This effect
was not taken into account in Ref. \cite{Bret1}. In addition, only
transverse plasma temperature was there accounted for, which
eventually does not happen to be restrictive.}. It is important to
note that the associated real parts $\Re(\omega_{\mathbf{k}})$ are
in the vicinity of the resonance given by $\omega-k_\parallel
V_{b}=0$. To our knowledge, this is the first exact computation of
TSF growth rates in the whole $k$-space including beam and plasma
temperatures effects. These curves clearly show that when
temperatures are accounted for they act to control the instability
domain, damping the small wavelengths perturbations along the
filamentation direction ($k_\parallel=0$) and deforming the growth
rates surface so that a maximum growth rate appears for a finite
oblique wave vector $\mathbf{k}_M$. In this respect, Fig. 2(b)
shows the drastic influence of beam transverse temperature for a
cold plasma. Yet, every physical plasma has a finite bulk
temperature and, for $\alpha$ small enough, this plasma
temperature can be shown \cite{Bret2} to control essentially the
maximum growth rate location. Its $(k_\perp,k_\parallel)$
components are then
\begin{equation}\label{eq:ZmaxRelat}
\mathbf{k}_M \sim \left(\pm\frac{\omega_p}{c}\sqrt{V_b/V_{tp}},\pm
\frac{\omega_p}{V_b}\right).
\end{equation}
We can then roughly evaluate the density perturbation in Eq.
(\ref{eq:n1fourier}) by only retaining the $\mathbf{k}_M$
contribution. As for the corresponding proper frequency, one has
$\omega_{k_M} \sim \pm \omega_p + i \delta_M$ where $\delta_M$ is
given by Eq. (\ref{eq:tauxmax}). Summing in the $\mathbf{k}$-space
the four contributions associated to all the possible orientations
of $\mathbf{k}_M$ in (\ref{eq:ZmaxRelat}), one finds that the
density perturbation behaves essentially as
\begin{equation}\label{eq:perturbation}
\rho_1(\mathbf{r},t)\propto \exp(\delta_M
t)\sin(k_{M\parallel}z-\omega_pt)\cos(k_{M\perp}x).
\end{equation}
Equation (\ref{eq:perturbation}) displays spacial modulation of
electron density in the beam direction ($z$) as well as in the
normal direction ($x$). In the normal direction, we witness the
``birth'' of the beam filamentation in the linear stage, with
filaments interspace
\begin{equation} \label{eq:filament_interspace}
L_f \sim \pi\lambda_s\sqrt{V_{tp}/V_b},
\end{equation}
where $\lambda_s=c/\omega_p$ is the skin depth.
\begin{figure}[t]
\begin{center}
\includegraphics[width=0.35\textwidth]{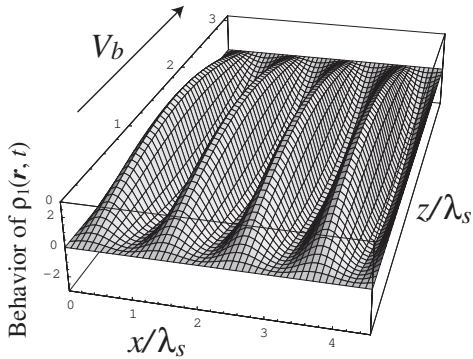}
\end{center}
\caption{Right hand side of Eq. (\ref{eq:perturbation}) for
$t=1/\omega_p$. Parameters are $V_{tp}=c/30$, $\alpha=0.1$ and
$\gamma_b=7$ yielding $\delta_M \sim 0.16\omega_p$.}
\label{fig:fila}
\end{figure}

There are by now very few relevant experimental results available
for quantitative comparisons with this result. We can consider
Fig. 3 of Ref. \cite{Tatarakis}, where plasma electronic density
is about 10$^{20}$ electrons/cm$^3$. This yields a plasma skin
depth of about 53 $\mu$m while Fig. 3 scale indicates the
transverse space between filaments is somehow smaller. Indeed, the
quantity $L_f$ introduced above is the skin depth times a
$\pi\sqrt{V_{tp}/V_b}$ factor which is smaller than 1 for a
non-relativistic plasma since $V_b \sim c$ here. Taking account of
the estimated plasma temperature (100 eV), we finally find $L_f
\sim$ 23 $\mu$m which is in good agreement with what is observed.
Figure \ref{fig:fila} displays the right hand side of Eq.
(\ref{eq:perturbation}) for $t=1/\omega_p$. Filaments are clearly
visible, combined with a beam segmentation along the beam
direction into segments $\pi\lambda_s V_b/c \sim \pi\lambda_s$
long. This parallel segmentation may not be easily distinguishable
on Fig. 3 of Ref. \cite{Tatarakis} for its characteristic length
(more than 150 $\mu$m) is comparable to the size of the entire
picture.

Let us here briefly discuss the applicability of our study. As far
as plasma (or beam) transverse temperatures are large enough, the
above results should apply to a finite system where the beam has a
finite radial extension $r_b$, the condition being that $r_{b}
k_{M\perp} \gg 1$ (see \cite{Ivanov} for filamentation instability
in a finite size beam). As far as FIS quantitative applications
are concerned, the major potential restriction of the present
study is the fact that the longitudinal beam temperature has been
neglected. Taking it into account would substantially increase its
difficulty as it would require a full kinetic treatment and may
render untractable the already demanding formal computations used
in Fig. \ref{fig:growthrates}. A useful discussion on the onset of
kinetic effects and breakdown of the cold beam hypothesis may be
found in Refs. \cite{Rudakov}. Besides, we used there waterbag
distributions which were simpler to tackle than Maxwellian, but
this should only affect marginally the quantitative results
obtained.

Let us summarize our point as a conclusion. Relativistic beam
filamentation is an observed phenomena. It is usually associated
with the exponential growth of an unstable mode called
``filamentation instability''. It turns out that a thorough study
of every unstable modes reveals that the ``filamentation mode''
should not be the most unstable. Furthermore, this mode is purely
transverse and therefore unable to produce charge density
perturbations. A better candidate to explain beam filamentation is
the most unstable mode all over the $\mathbf{k}$ space, which
turns to be intermediate between filamentation and two-stream
waves. Not only this mode appears to be the fastest growing one,
it is also quasi-longitudinal so that it can perfectly induce
charge density perturbations. A simple evaluation of its growth
shows how it creates beam filaments within a few plasma periods
and agreement with experiment presented in \cite{Tatarakis} is
found to be correct. It sets the characteristic transverse and
longitudinal filamentation scales, at least during the linear
initial stage when resistive (collisional) effects are still
negligible \cite{Ivanov}. Finally, we wish to mention that our
study emphasizes the importance of quasi-longitudinal modes in
modelling filamentation which agrees with some considerations
recently put forward by Macchi {\it{et al.}} \cite{Macchi} among
others.

\bibliographystyle{unsrt}
\bibliography{BibBret}

\begin{thebibliography}{10}

\bibitem{Tabak}
M.~Tabak, J.~Hammer, M.~E. Glinsky, W.~L. Kruer, S.~C. Wilks, J.~Woodworth,
  E.~M. Campbell, M.~D. Perry, and R.~J. Mason.
\newblock {\em Phys. Plasmas}, 1:1626, 1994.

\bibitem{Ruhl}
H.~Ruhl, A.~Macchi, P.~Mulser, F.~Cornolti, and S.~Hain.
\newblock {\em Phys. Rev. Lett.}, 82:2095, 1999.

\bibitem{Dodd}
E.~S. Dodd, R.~G. Hemker, C.-K. Huang, S.~Wang, C.~Ren, W.~B. Mori, S.~Lee, and
  T.~Katsouleas.
\newblock {\em Phys. Rev. Lett.}, 88:125001, 2002.

\bibitem{silva2002}
L.~O. Silva, R.~A. Fonseca, J.~W. Tonge, W.~B. Mori, and J.~M. Dawson.
\newblock {\em Phys. Plasmas}, 9:2458, 2003.

\bibitem{Fonseca2003}
R.~A. Fonseca, L.~O. Silva, J.~W. Tonge, W.~B. Mori, and J.~M. Dawson.
\newblock {\em Phys. Plasmas}, 10:1979, 2003.

\bibitem{Sentoku}
Y.~Sentoku, K.~Mima, P.~Kaw, and K.~Nishikawa.
\newblock {\em Phys. Rev. Lett.}, 90, 2003.

\bibitem{Tatarakis}
M.~Tatarakis, F.~N. Beg, E.~L. Clark, A.~E. Dangor, R.~D. Edwards, R.~G. Evans,
  T.~J. Goldsack, K.~W.~D. Ledingham, P.~A. Norreys, M.~A. Sinclair, M-S. Wei,
  M.~Zepf, and K.~Krushelnick.
\newblock {\em Phys. Rev. Lett.}, 90:175001, 2003.

\bibitem{LeeLampe}
R.~Lee and M.~Lampe.
\newblock {\em Phys. Rev. Lett.}, 31:1390, 1973.

\bibitem{Hubbard}
R.F. Hubbard and D.A. Tidman.
\newblock {\em Phys. Rev. Lett.}, 41:866, 1978.

\bibitem{Okada1980}
T.~Okada and K.~Niu.
\newblock {\em J. Plasma Phys.}, 24:483, 1980.

\bibitem{Okadacollisions}
T.~Okada and W.~Schmidt.
\newblock {\em J. Plasma Phys.}, 37:373, 1987.

\bibitem{Molvig}
K.~Molvig.
\newblock {\em Phys. Rev. Lett.}, 35:1504, 1975.

\bibitem{Weibel}
E.~S. Weibel.
\newblock {\em Phys. Rev. Lett.}, 2:83, 1959.

\bibitem{Bret1}
A.~Bret, M.C. Firpo, and C.~Deutsch.
\newblock {\em Phys. Rev. E}, 70:046401, 2004.

\bibitem{Fainberg}
Ya.~B. Fa\u{\i}nberg, V.D. Shapiro, and V.I. Shevchenko.
\newblock {\em Soviet Phys. JETP}, 30:528, 1970.

\bibitem{Bret2}
A.~Bret, M.C. Firpo, and C.~Deutsch.
\newblock {\em unpublished}.

\bibitem{califano3}
F.~Califano~R. Prandi, F.~Pegoraro, and S.~V. Bulanov.
\newblock {\em Phys. Rev. E}, 58:7837, 1998.

\bibitem{Ivanov}
A.~A. Ivanov and L.I. Rudakov.
\newblock {\em Soviet Phys. JETP}, 58:1332, 1970.

\bibitem{Rudakov}
L.I. Rudakov.
\newblock {\em Soviet Phys. JETP}, 59:209, 1970.

\bibitem{Macchi}
A.~Macchi, A.~Antonicci, S.~Atzeni, D.~Batani, F.~Califano, F.~Cornolti, J.~J.
  Honrubia, T.~V. Lisseikina, F.~Pegoraro, and M.~Temporal.
\newblock {\em Nucl. Fusion}, 43:362, 2003.

\end{thebibliography}

\end{document}